\documentclass[prl,twocolumn,floatfix,nobibnotes,superscriptaddress,prb]{revtex4}
\usepackage{amsmath}
\usepackage{graphicx}

\begin{document}
\title{Anomalous thermal expansion in $\alpha$-titanium}

\author{P. Souvatzis}
\affiliation{Department of Physics, Uppsala University,
Box 530, SE-75121, Uppsala, Sweden}
\author{O. Eriksson}
\affiliation{Department of Physics, Uppsala University,
Box 530, SE-75121, Uppsala, Sweden}
\author{M. I. Katsnelson}
\affiliation{Institute for Molecules and Materials, Radboud
University Nijmegen, NL-6525 ED, Nijmegen, The Netherlands}
\date{\today }

\begin{abstract}
We provide a complete quantitative explanation for the anisotropic
thermal expansion of hcp Ti at low temperature. The observed
negative thermal expansion along the c-axis is reproduced
theoretically by means of a parameter free theory which involves
both the electron and phonon contributions to the free energy. The
thermal expansion of titanium is calculated and found to be
negative along the c-axis for temperatures below $\sim$ 170 K, in
good agreement with observations. We have identified a
saddle-point Van Hove singularity near the Fermi level as the main
reason for the anisotropic thermal expansion in $\alpha-$titanium.


\end{abstract}
\maketitle

The most general aspects of the chemical bonding in the transition
$d$ metals can be understood from the Friedel
model\cite{harrison}, explaining the trends in equilibrium volume,
bulk modulus and cohesive energy. The transition metals are found
to crystallize at low temperatures in the cubic fcc and bcc
structures, and the hexagonal hcp structure~\cite{footnote1},
which can be qualitatively explained from a band filling of
itinerant {\it d}-states\cite{skriver}. In addition, the Debye
model reproduces the thermal volume expansion with a rather good
accuracy\cite{moruzzi}. Hence, with a seemingly good understanding
of the fundamental mechanisms governing the properties of the
transition metals, the recently observed negative thermal
expansion coefficient along the c-axis of one of these elements,
the hcp ($\alpha$) phase of Ti~\cite{Misha}, stands out as an
enigma. Especially since no other transition metal so far has been
shown to display such a behaviour.

The problem of finding connections between the electronic
structure of metals and alloys and peculiarities of their lattice
properties has a long history, starting with the ``third
Hume-Rothery rule'' concerning boundaries of phase stability in
noble-metal alloys~\cite{hume} and its explanation by Jones in
terms of touching of the Brillouin zone faces by the Fermi
sphere~\cite{jones} (for a review of further developments of these
ideas, see Ref.~\cite{KNT}). The general concept of electronic
topological transitions (ETT), introduced by I.
Lifshitz~\cite{Lif}, that is, a coincidence of the Fermi level
with a Van Hove singularity of the electronic density of states
(DOS), is of crucial importance for understanding these
interrelations. Phase transitions and pre-martensitic anomalies of
elastic moduli in alkali and alkaline-earth metals under pressure
provide a clear example of the effects of the Van Hove
singularities on the lattice properties~\cite{vaks1,vaks2}. It
turns out that the singularity in the electron DOS at the Fermi
energy, $N(E_F)$, should be visible also in elastic moduli and
Debye temperature and, thus, in the thermodynamic properties of
metals at low enough temperatures (the anomalies in phonon spectra
with large enough wave vectors and thus in high-temperature
thermodynamic properties are in general weaker, see
Ref.~\cite{KNT} and references therein). Since the thermal
expansion is connected with the pressure derivatives of the
elastic moduli, anomalies in the thermal expansion might be
especially strong. In particular, it can be proven
thermodynamically that ETT in non-cubic metals should lead to a
singular anisotropic thermal expansion at low enough
temperatures~\cite{Misha,Antropov}. The latter means that in
principle it is always possible to prepare a textured material
with zero thermal expansion. This conclusion, being interesting in
itself, opens new ways to find nonmagnetic Invar systems. However,
based on these general considerations alone it is impossible to
predict the temperature region where the effect should be
observable, or how far from the point of ETT the effect is still
visible. Here we answer these questions based on direct
microscopic calculations, in a framework of the density functional
theory, and we address the recently discovered negative thermal
expansion of $\alpha$-Ti~\cite{Misha}.


The occurrence of negative thermal expansion at low temperatures
for non-cubic elemental solids have been known for quite some
time, but only for elements outside the transition metal series.
For instance, the hexagonal close packed metals zinc and
cadmium~\cite{Munn} have negative thermal expansion coefficients
along the a-axis ($\alpha_{\bot}$) for temperatures below $\sim$ 75
K, while amongst the IIIB group of the Periodic Table it is tin
and indium that have negative thermal expansion coefficients along
the basal plane ($\alpha_{\bot}$) and orthogonal to the basal
plane ($\alpha_{||}$), respectively~\cite{Munn}. Amongst the
transition metals, Ti however stands out.



The analysis presented here is based on first principles density
functional theory of the electron and phonon contributions to the
total energy. We write the Helmholtz free energy as
\begin{eqnarray}\label{eq:free}
F(\bar{\epsilon},T) =  \qquad \qquad \qquad \qquad \qquad \qquad \qquad \qquad
\nonumber \\ \frac{V}{2}\sum_{ij} C_{ij}\epsilon_{i}\epsilon_{j}+
F^{phon}(\bar{\epsilon},T)+F^{el}(\bar{\epsilon},T),
\end{eqnarray}
where $C_{ij}$ are the elastic constants, $\bar{\epsilon}$ the
elastic strain, $V$ the volume, $F^{phon}$ the phonon free energy
and $F^{el}$ the energy of thermal excitations in the electron
subsystem. In this expression the reference (zero) level is for a
crystal at equilibrium conditions at zero temperature. The elastic
constants were calculated from first principles\cite{method}.

To evaluate the free energy contribution $F^{phon}$, which can be
expressed as~\cite{wallace,wille}
\begin{eqnarray}\label{eq:phonF}
F^{phon}(\bar{\epsilon},T) = \qquad \qquad \qquad \qquad \qquad \qquad  \qquad \nonumber \\
\int_{0}^{\infty}d\omega g(\omega,\bar{\epsilon})[\frac{\hbar\omega}{2}+k_{B}Tln(1-e^{-\hbar\omega/k_{B}T})],
\end{eqnarray}
the phonon DOS $g(\omega,\bar{\epsilon})$ has to be calculated.
This was done within the quasi-harmonic approximation \cite{YUN,PETROS}, where
all anharmonic effects except the thermal expansion 
are neglected when calculating the temperature dependence of the phonons.
In practice the phonon DOS was calculated
by making small displacements of the atoms in a supercell
(SC)~\cite{PETROS}. The directions of the displacements were $[1 1
0]$ and $[0 0 1]$  with amplitudes that were equal to $\sim$0.4\%
of the lattice constant. The supercell used was a 3x3x2 cell.
Further details are found in Ref.\onlinecite{method}. 
The energy of thermal excitations of electron states $F^{el}$ was
calculated by the standard expression by Sommerfeld and Frank \cite{Sommer}.
\begin{equation}\label{eq:freesom}
F^{el}(\bar{\epsilon},T) = -\frac{(\pi k_{B})^2}{6}D(\epsilon_{F},\bar{\epsilon})T^{2}.
\end{equation}
where $D(\epsilon_{F},\bar{\epsilon})$ is the calculated electronic density of states at the Fermi level.

In Fig.1 we compare
our calculated phonon spectrum (at $T=0$) with
experimental values (at room temperature). It is worthwhile to
mention that a tight-binding calculation of the phonon dispersion
for hcp Ti has been published recently~\cite{sven-ti}, where
parameters of the model were fitted to experimental data as well
as to first principles calculations. The theoretical phonon
dispersion curve in Fig.1 agrees very well with the theoretical
curves in Ref.~\cite{sven-ti}. When comparing the theoretical and
experimental~\cite{Stassis} curves, we note an overall agreement,
although certain differences can be identified. For instance,
along the $\Gamma-A$ direction the theory underestimates the
frequencies in the lowest experimental branch, whereas the higher
branches are reproduced with better accuracy, especially using
the general gradient approximation (GGA). Also, the calculated lowest branch along the $\Gamma-K$
direction comes out somewhat too low compared to observations. The
phonon DOS was then calculated with the method of Ref.~\cite{Alfe}.
\begin{figure}[tbp]
\begin{center}
\includegraphics*[angle=0,scale=0.35]{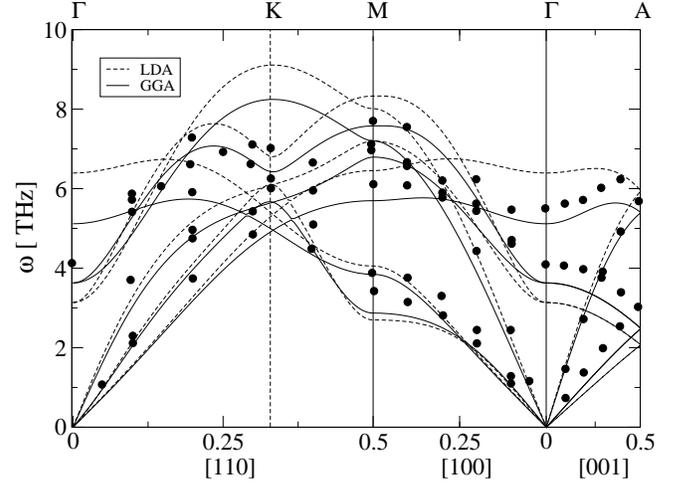}
\caption{The phonon dispersion of hcp Ti at room temperature and
ambient pressure. The solid curve is the calculated frequencies
from GGA and the dashed curve represents calculated frequencies
from LDA. Both calculations are done at the experimental volume
$V_{0} = 15.91$\AA$^3$.
The filled circles are the experimental data of Stassis $et$ $al$
\cite{Stassis}.} \label{fig:freq}
\end{center}
\end{figure}



By differentiating the free energy (\ref{eq:free}) with respect to
$\epsilon_{v}\equiv d(ln(V))$ and $\epsilon_{c} \equiv d(ln(c/a))$, it is
possible to obtain an expression for the change in volume and structural property as
a function of temperature. These changes are expressed in terms of equilibrium strains
,$\epsilon_{v}^{0}$ and $\epsilon_{c}^{0}$ at which $\partial F/\partial \epsilon_{v}=\partial F/\partial \epsilon_{c} = 0$,
 and can be written in terms of the
elastic constants and strain derivatives of the free energy
\begin{eqnarray}
\epsilon_{v}^{0}(T) = \frac{1}{V(B_{11}B_{22}-B_{12}^{2})}\Big
[-B_{22}\frac{\partial F^{*}}{\partial \epsilon_{v}}
 + B_{12}\frac{\partial F^{*}}{\partial \epsilon_{c}}
\Big ] \label{eq:strain1} \\
\epsilon_{c}^{0}(T) = \frac{1}{V(B_{11}B_{22}-B_{12}^{2})}\Big
[B_{12}\frac{\partial F^{*}}{\partial \epsilon_{v}}
 \quad - B_{11}\frac{\partial F^{*}}{\partial \epsilon_{c}}
\Big ]\label{eq:strain2}
\end{eqnarray}
where
\begin{eqnarray}
B_{11} &=& \frac{2}{9}( C_{11}+C_{12}+\frac{1}{2}C_{33}+2C_{13} ) \label{eq:elastic1} \\
B_{22} &=&  \frac{2}{9}( C_{11}+C_{12}+2C_{33}-4C_{13} ) \label{eq:elastic2} \\
B_{12} &=& \frac{1}{9}( C_{33}+C_{13}-C_{11}-C_{12} )  \label{eq:elastic3} \\
F^{*} &=&  F^{phon} + F^{el}. \label{eq:freetot}
\end{eqnarray}
Furthermore by differentiating (\ref{eq:strain1}) and
(\ref{eq:strain2}) with respect to the temperature the following
relations are obtained for the thermal expansion
coefficients~\cite{Misha,Antropov}
\begin{eqnarray}
\alpha_{a} = \frac{1}{3V(B_{11}B_{22}-B_{12}^2)}\Big [-(B_{22}+B_{12})\frac{\partial^{2}F^{*}}{\partial T \partial \epsilon_{v}}
\qquad \quad \nonumber \\
+(B_{12}+B_{11})\frac{\partial^{2}F^{*}}{\partial T \partial \epsilon_{c}}
 \Big ] \qquad \label{eq:inpl} \\
\alpha_{c} = \frac{1}{3V(B_{11}B_{22}-B_{12}^2)}\Big [-(B_{22}-2B_{12})\frac{\partial^{2}F^{*}}{\partial T \partial \epsilon_{v}}
\qquad \nonumber \\
+(B_{12}-2B_{11})\frac{\partial^{2}F^{*}}{\partial T \partial \epsilon_{c}}
 \Big ] \qquad \label{eq:outpl} \\
\beta = \frac{1}{V(B_{11}B_{22}-B_{12}^{2})}\Big [-B_{22}\frac{\partial^{2} F^{*}}{\partial T \partial \epsilon_{v}}
 + B_{12}\frac{\partial^{2} F^{*}}{\partial T
\partial \epsilon_{c}}\Big ], \qquad \label{eq:vol}
\end{eqnarray}
where $\alpha_{a}=\frac{1}{a}\frac{da}{dT}$, $\alpha_{c}=\frac{1}{c}\frac{dc}{dT}$ and
$\beta=\frac{1}{V}\frac{dV}{dT}$.
\newline

By fitting free energies calculated at different strains
and at
a given temperature to polynomials of first degree in
$\epsilon_{v}$ and second degree in $\epsilon_{c}$, the
equilibrium strains can be obtained from Eqns. (\ref{eq:strain1})
and (\ref{eq:strain2}), and the thermal expansion coefficients can
be calculated from Eqns. (\ref{eq:inpl})- (\ref{eq:vol}).

\begin{figure}[tbp]
\begin{center}
\includegraphics*[angle=0,scale=0.35]{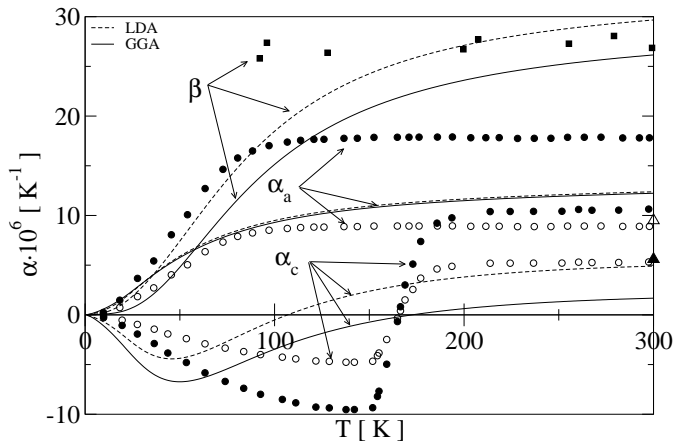}
\caption{Linear thermal expansion for hcp Ti at ambient pressure. The
solid lines are the theoretical calculation using GGA and the dashed line is from an LDA calculation. The filled circles are the experimental data of Nizhankovskii
 $et$ $al$ \cite{Misha}. The filled squares are the experimental volume expansion coefficient of Mal'ko $et$ $al$ \cite{Malko}.
The filled and empty triangle are the experimental data of $\alpha_{a}$ and $\alpha_{c}$ respectively
of Pawar $et$ $al$ \cite{Ram}. The empty circles are the experimental data of Nizhankovskii  $et$ $al$ scaled
to give a volume expansion coefficient in agreement with Ref.\onlinecite{Malko}.}
\label{fig:linexp}
\end{center}
\end{figure}

In Fig.\ref{fig:linexp} we show the calculated thermal expansion
coefficients of $\alpha$-titanium. The most important information
to be extracted from this figure is that the observed negative
thermal expansion coefficient along the c-axis is reproduced by
our theory, where especially the calculation based on GGA
reproduce observations with the highest accuracy. It should be noted
that GGA often is found to describe chemical bonding with better accuracy
than LDA. The temperature interval for which $\alpha_{c}$ is
negative is roughly 0-170 K, both in the observations and from the
theory. The order of magnitude of  $\alpha_{c}$ and
$\alpha_{a}$ is also the same when comparing experiment and theory.
Figure \ref{fig:linexp} also shows that
theory reproduces, with good accuracy, the volume expansion
coefficients of Ref.~\onlinecite{Malko}, especially at
somewhat elevated temperatures. We also note that based on thermodynamic
relations $\beta$ should approach zero at T = 0 K, which our theoretical curves do.

%


\bigskip
\bigskip

The fact that both the measured and calculated thermal expansion
coefficients along the c-axis of Ti are negative at low temperatures
strongly suggests the uniqueness of elemental Ti among transition metals,
 although the absolute value of the measures low temperature expansion coefficient
is still somewhat uncertain. The measured data of Ref. \onlinecite{Misha} (filled circles) have
in Fig. \ref{fig:linexp} been scaled (open circles) to reproduce room temperature
values of $\beta$, $\alpha_{a}$ and $\alpha_{c}$, and it is found that these scaled
values compare better with our theory (Fig. \ref{fig:linexp}). Although a slight
calibration error in Ref. \onlinecite{Misha} can not be excluded, there is good
reason to view the negative value of $\alpha_{c}$ at low temperatures as a true
materials property of $\alpha$-Ti.

\begin{figure}[tbp]
\begin{center}
\includegraphics*[angle=0,scale=0.41]{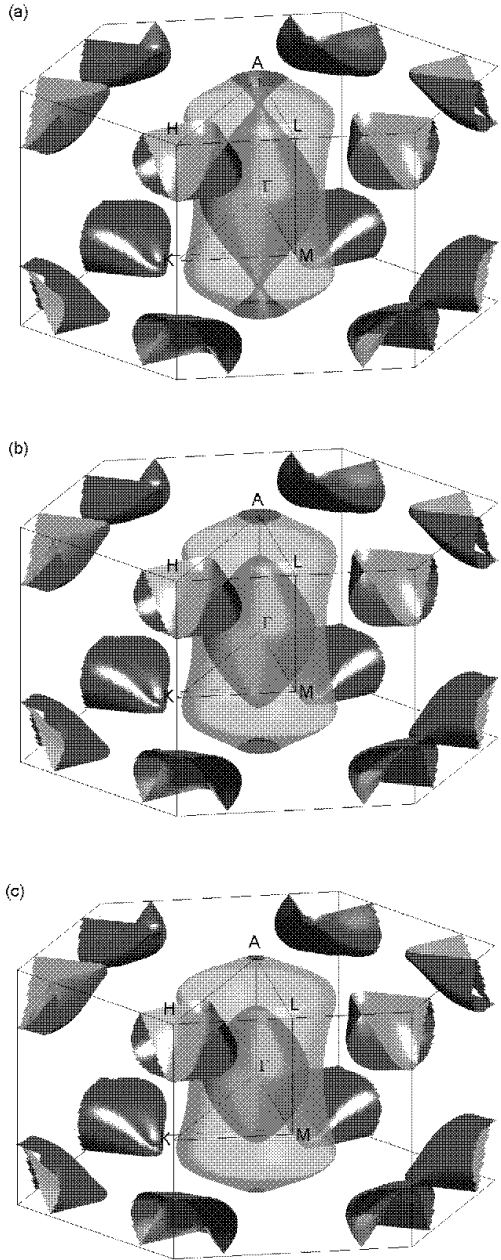}
\caption{ Calculated Fermi surface of hcp Ti at T=0 equilibrium volume for three different lattice
constants c, at $c = 0.988c_{0}$ (a), at $c=c_{0}$ (b) and at $c = 1.012c_{0}$ (c). Here
$c_{0}$ corresponds to the (T=0) equilibrium lattice constant.}
\label{fig:fermis}
\end{center}
\end{figure}

As we will show below the microscopic origin of the negative thermal expansion for $\alpha_{c}$ of Ti is due to the closeness to a saddle point van Hoove singularity of the electronic structure. 
To illustrate this singularity we proceed with an analysis of the Fermi surface. 
In order to do this we show in Fig.
\ref{fig:fermis} the calculated Fermi surface at the equilibrium
volume for three different values of the out-of-plane lattice
constant, c. The figure shows that as the c
lattice constant decreases
the inner ellipsoidal surface at the $\Gamma$-point and the Fermi surface centered at the
$A$-point, become connected along the $\Gamma A$ line.
The electronic structure as revealed by the Fermi surface shown in Fig.3 thus demonstrates the presence of a
saddle point Van
Hove singularity, which is associated with a singular
contribution to the density of states at the Fermi level
$N(E_{F}),$ $\delta N \left( E_F \right)
\sim-\sqrt{E_{F}-E_{c1}}\theta(E_{F}-E_{c1})$ , where $\theta(x)$
fulfills: $\theta(x>0) = 1$, $\theta(x<0) = 0$ and $E_{c1}$ is the
critical point energy \cite{Misha,Antropov}. The energy difference
between $E_F$ and the energy of the
critical saddle point, at the theoretical equilibrium volume and a c=$c_0$, has
been calculated to be $E_{F}-E_{c1}\sim 44$ meV.
Another critical point, associated
with the appearance of a new ellipsoid around the $K$ symmetry
point (not shown in  Fig. \ref{fig:fermis} ) has been found in the calculations,
giving rise to the singular contribution $\delta N(E_{F}) \sim
\sqrt{E_{c2}-E_{F}}\theta(E_{c2}-E_{F})$ to
$N(E_{F})$. However since $E_{c2}-E_{F}\sim
142$ meV  and $|\partial (E_{c2}-E_{F})/\partial \epsilon | \ll |
\partial (E_{F}-E_{c1})/\partial \epsilon |$,
 it is clear that the saddle-point topological transition, at $E_{c1}$,
gives rise to the strongest singular contribution to $N(E_{F})$.

By calculating the derivatives of $N(E_{F})$ with respect to the two
different types of strains we have found that $\partial
N(E_{F})/\partial \epsilon_{v} \sim 0.75$ eV$^{-1}$ and $\partial
N(E_{F})/\partial \epsilon_{c} \sim -0.77$ eV$^{-1}$. Since
the singularities in $N(E_{F})$ influence the
elastic moduli, thus effecting the Debye temperature \cite{KNT}, it is natural to attribute
the main reason for the anisotropic thermal expansion in $\alpha-$titanium to the
saddle-point Van Hove singularity near the Fermi level.

{\bf Acknowledgments} We are grateful to the Strategic Foundation
for Research (SSF), the Swedish Research Council (VR), the Swedish
National Supercomputer Center (NSC), UPMAX and to the G\"oran
Gustafsson foundation, for support. MIK acknowledges a support
from Stichting voor Fundamenteel Onderzoek der Materie (FOM), the
Netherlands. Valuable discussions with Prof. U. Jansson are
acknowledged.



\begin{references}

\bibitem{harrison} W. A. Harrison, {\it Electronic Structure and the Properties of
Solids} (W.H.Freeman and Company, San Francisco, 1980).

\bibitem{footnote1} With the exception of the complex structure of Mn at low temperatures.

\bibitem{skriver} H. L. Skriver, Phys. Rev. B {\bf 31}, 1909 (1985).

\bibitem{moruzzi} V. L. Moruzzi,  J. F. Janak,  \& K. Schwarz, Phys. Rev. B {\bf 37}, 790 (1988).

\bibitem{Misha} V. I. Nizhankovskii, M. I. Katsnelson, G. V. Peschanskikh,  \& A. V.  Trefilov, Pis'ma ZhETF {\bf 59},
693 (1994).

\bibitem{hume} W. Hume-Rothery, {\it The Metallic State} (Oxford Univ. Press, 1931)

\bibitem{jones} N. F. Mott, \& H. Jones, {\it The Theory of the Properties of Metals and Alloys} (Oxford Univ. Press, 1936).

\bibitem{KNT} M. I. Katsnelson, I.I. Naumov, I.I. \& A. V. Trefilov, Phase Transitions {\bf 49}, 143 (1994).

\bibitem{Lif} I. M. Lifshitz, Sov. Phys. JETP {\bf 11}, 1130 (1960).

\bibitem{vaks1} V. G. Vaks {\it et al.}, J. Phys.: Condens. Matter {\bf 1}, 5319 (1989).

\bibitem{vaks2} V. G. Vaks {\it et al.} J. Phys.: Condens. Matter {\bf 3} 1409 (1991).








\bibitem{Antropov} V. P. Antropov {\it et. al.}, Phys.
Lett. A {\bf 130}, 155 (1988).


\bibitem{Munn} R. W. Munn, {\it The Thermal Expansion of Axial Metals}, Advances in physics,  {\bf 18 } (1969).

\bibitem{method} The first principles calculations were done with
the VASP code~\cite{vasp}. Convergence in sampling of the
Brillouin zone was obtained at  20480 k-points (for elastic
constants) and 486 k-point (for phonon calculations). The
calculations employed both the local density approximation (LDA) and the
general gradient approximation (GGA).
The phonon density of states were calculated
using a $150\times 150\times 100$ k-point mesh with a
0.05 THz smearing. The calculations have been performed for 15
different volume strains $\epsilon_{v}$ in the
range 0 $\le \epsilon_{v} \le 0.013$, and at each volume strain
for three different tetragonal strains $\epsilon_{c} =
\epsilon^{0}_{c}-0.001,\epsilon^{0}_{c},\epsilon^{0}_{c}+0.001$,
where $\epsilon^{0}_{c}$ is the tetragonal strain corresponding to
the minimum static lattice energy at a given volume strain, $\epsilon_{v}$.
In the VASP calculations a cut-off of 302 eV was used. To each
eigenvalue a Gaussian smearing of $\sim$ 0.2 eV was applied to
speed up the convergence of the calculation.

\bibitem{vasp} G. Kresse \& J. Furthmuller,  Phys. Rev. B {\bf 54}, 11169 (1996).

\bibitem{wallace} D. C. Wallace, {\it Thermodynamics of Crystals} ( Wiley, New York, 1972 ).

\bibitem{wille} G. K. Straub, J. B. Aidun, J. M. Wills, C. R. SanchezCastro, \& D. C. Wallace,
Phys. Rev. B {\bf 50}, 5055 (1994).

\bibitem{YUN}Yu. N. Gornostyrev et al, Scripta Metal. {\bf 56}, 81 (2007)

\bibitem{PETROS} P. Souvatzis, A. Delin, \& O. Eriksson,
 Phys. Rev. B {\bf 73}, 054110 (2006).

\bibitem{Sommer} A. Sommerfeld \& N. H. Frank, Rev. Mod. Phys. {\bf 3},
1 (1931) .

\bibitem{sven-ti} D. R. Trinkle {\it et. al.}, Phys. Rev. B {\bf 73}, 94123 (2006).

\bibitem{Stassis} C. Stassis, D. Arch, B. N. Harmon \& N. Wakabayashi, Phys. Rev. B {\bf 19}, 181 (1979).

\bibitem{Alfe} D. Alf\'e, The {\it Phon}-software together with a description
of the program  can be found at:
http://chianti.geol.ucl.ac.uk/~dario/ .




\bibitem{Malko} P. I. Mal'ko, D. S. Arensburger, V. S. Pugin, V. F. Nemchenko, \&  S. N. L'Vov,
Powder Metallurgy and Metal Ceramics, {\bf 9}, 642 ( 1970).

\bibitem{Ram} R.R Pawar, and V.T. Deshpande, Acta Crystallogr. A {\bf 24}, 316 (1968).





\end{references}
\end{document}